\documentclass[]{raa}            
\usepackage{graphicx,times}
\usepackage{subcaption}
\usepackage{caption}
\usepackage{amssymb}
\usepackage{amsmath}
\usepackage{natbib}
\begin{document}
   \title{Mode identification revisit and asteroseismology of the DAV star TIC 231277791
   
}

\volnopage{ {\bf 0000} Vol.\ {\bf 0} No. {\bf XX}, 000--000}
\setcounter{page}{1}

\author{Z. K. Yao\inst{1,2,3}$^*$ \ Y. H. Chen\inst{1,2,3}$^*$ \ M. Y. Tang\inst{2}}

\institute{\inst{1} Faculty of Science, Kunming University of Science and Technology, Kunming 650093, China; {$1931824997@qq.com$}\\
           \inst{2} Institute of Astrophysics, Chuxiong Normal University, Chuxiong 675000, China; {$yanhuichen1987@126.com$}\\
           \inst{3} International Centre of Supernovae (ICESUN), Yunnan Key Laboratory, Kunming 650216, China{}
\\
\vs \no
{\small Received [0000] [July] [day]; accepted [0000] [month] [day] }}

\abstract{White dwarfs are the final stage for most low and intermediate mass stars, which plays an important role in understanding stellar evolution and galactic history. Here we performed an asteroseismological analysis on TIC 231277791 based on 10 independent modes reported by Romero et al. Two groups of modes were identified with frequency splitting: mode identification$\_{1}$ with one $l$\,=\,1, $m$\,=\,0 mode, two $l$\,=\,2, $m$\,=\,0 modes, and three $l$\,=\,1 or 2, $m$\,=\,0 modes, and mode identification$\_{2}$ wtih one $l$\,=\,1, $m$\,=\,0 mode, three $l$\,=\,2, $m$\,=\,0 modes, and one $l$\,=\,1 or 2, $m$\,=\,0 mode. The rotation period is derived to be 41.64\,$\pm$\,2.73\,h for TIC 231277791. We established a large sample (7,558,272) of DAV star models using the White Dwarf Evolution Code (\texttt{WDEC}; 2018, v16), resulting of optimal models with model$\_{1}$ (mode identification$\_{1}$):  $M_\mathrm{*}$\,=\,0.570\,$\pm$\,0.005\,$M_\mathrm{\odot}$, $T_\mathrm{eff}$\,=\,11300\,$\pm$\,10\,K, -log($M_\mathrm{H}/M_\mathrm{*}$)\,=\,9.15\,$\pm$\,0.01, -log($M_\mathrm{He}/M_\mathrm{*}$)\,=\,4.94\,$\pm$\,0.01, and $\sigma_{\textup{RMS}}$\,=\,0.06\,s, and model$\_{2}$ (mode identification$\_{2}$): $M_\mathrm{*}$\,=\,0.720\,$\pm$\,0.005\,$M_\mathrm{\odot}$, $T_\mathrm{eff}$\,=\,11910\,$\pm$\,10\,K, -log($M_\mathrm{H}/M_\mathrm{*}$)\,=\,6.11\,$\pm$\,0.01, -log($M_\mathrm{He}/M_\mathrm{*}$)\,=\,3.09\,$\pm$\,0.01, and $\sigma_{\textup{RMS}}$\,=\,0.04\,s. The central oxygen abundances are 0.71 (optimal model$\_{1}$) and 0.72 (optimal model$\_{2}$), respectively, which are consistent with the results of stellar structure and evolution theory.
 }

\authorrunning{Z. K. Yao, Y. H. Chen, M. Y. Tang}       
\titlerunning{Asteroseismology on TIC 231277791}  
\maketitle

\section{Introduction}

White dwarfs (WDs) are the final stage of evolution for the majority of stars in the universe (Saumon et al. 2022). They are composed of electron degenerate cores and ideal gas atmospheres, which plays an important role for understanding stellar evolution and galactic history. The spectral type of a WD is determined by its atmospheric chemical compositions. Kepler et al. (2021) reported that $\sim$80\% of WDs have hydrogen (H)-rich atmospheres (spectral class DA) and $\sim$20\% of WDs have helium (He)-rich atmospheres. The latter are called DB and DO WDs as determined by neutral He-rich or ionized He-rich atmospheres, respectively.
As stopped fusion in the core, the evolution of WDs is mainly dominated by cooling together with contraction (Althaus et al. 2010a). The cooling rate  is affected by parameters such as effective temperature ($T_\mathrm{eff}$), mass ($M_\mathrm{*}$), core chemistry, and atmosphere chemistry (Winget \& Kepler 2008). 

The first variable WD was discovered by Arlo Landolt, with a dominant period of $\sim$\,750\,s (Landolt 1968). The variable WDs exhibit non-radial $g$-mode pulsations with multi-periodic brightness variations. The combination of the $\kappa$-$\gamma$ mechanism (Dolez \& Vauclair 1981; Winget et al. 1982) and convective driving mechanism (Brickhill 1991; Goldreich \& Wu 1999) is the reason for the excitation of pulsations in WDs. Variable DA WDs, also known as ZZ Ceti stars or DAVs, are the most common class of pulsating WDs. To date, more than 500 DAVs have been confirmed (Romero et al. 2024). The pulsating instability strip for DAVs ranges roughly from $T_\mathrm{eff}$\,=\,12,270 to 10,850\,K (Gianninas, Bergeron, \& Ruiz 2011). The DAVs have pulsation periods basically from 100 to 1400\,s and pulsation amplitudes basically from 0.01 to 0.3\,mag (C$\acute{o}$rsico et al. 2019).

The internal structure of pulsating WD stars can be probed with the technique of asteroseismology (see, e.g., C$\acute{o}$rsico et al. 2019 and references therein). In this front, it has been seen that rapid progress based on photometric observations through single-station observations, multi-station joint observations, the WET runs (Nather 1990). Nowadays, photometry from space such as the Kepler Telescope (Kepler; Borucki et al. 2010), the K2 mission (Howell et al. 2014), and the Transiting Exoplanet Survey Satellite (TESS; Ricker et al. 2015) have revolutionized the WD asteroseismology in two ways (C$\acute{o}$rsico 2020, 2022). Firstly, it provides a more accurate measurement of pulsation periods than before, for instance, TESS has achieved $10^{-4}$\,s accuracy (Giammichele et al. 2022). Secondly, they provided many new pulsating WDs. Romero et al. (2022, 2024) found 106 new DAVs using TESS data from the first five years (Sectors 1-69). Hermes et al. (2017) reported a mean rotation period of 35\,$\pm$\,28\,h, for 0.51-0.73\,$M_\mathrm{\odot}$ WDs from K2 photometry. The long and continuous photometry also reveals a new and interesting phenomenon in pulsating white dwarfs as documented in Zong et al. (2016) who found amplitude modulation of pulsation modes. This feature can be helpful for mode identification, which established a profound foundation for the unprecedented precise chemical profile in pulsating WD stars (Giammichele et al. 2018). It leads a topic of debate in
the community about the abundance of carbon (C) and oxygen (O) in WD central cores (Giammichele et al. 2018, 2022).

Currently, there are mainly three methods to construct grid models of WDs. The first method constructs pure static WD models with parameterised chemical profiles (Giammichele et al. 2018, 2022). The second method adopts the fully evolutionary models of WDs, which calculates stellar evolutions from the zero-age main-sequence stage to the target WD stage (Paxton et al. 2011, Althaus et al. 2010b). The third method only calculates the evolution process of WDs, which are called quasi-static WD models (Bischoff-Kim \& Montgomery in 2018).

In 2022, Romero et al. reported 74 new bright DAVs, possibly includes 13 low-mass and one extremely low-mass WD candidate from TESS Sectors 1–39. TIC 231277791 ({RA=02:49:18.23, DEC=--53:34:35.4}) is one of the target samples with an observation period of 120\,s in sector 29 and 20\,s in sector 30. Subsequently, Romero et al. (2022) conducted spectroscopic observations for 29 objects, including the TIC 231277791, to refine atmospheric parameters and confirm the spectral characteristics of DA WDs. Finally, TIC 231277791 was identified as a DAV star. 
According to Romero et al. (2022), TIC 231277791 has 10 independent modes within a very small period range (from 497\,s to 768\,s), which should have obvious frequency splitting effect and opens up a possibility of mode identifications and model fittings. 
In this work, we first conducted detailed mode identifications on TIC 231277791. Then we evolved a large sample of DAV star models containing 7.56 million DAVs. Finally, a detailed asteroseismological model fitting was carried out on the DAV star TIC 231277791. The structure of this article is as follows. In Section 2, we introduce the mode identifications of TIC 231277791 and the model calculations. An asteroseismological study of TIC 231277791 is performed in Section 3. At last, we provide a discussion and conclusions in Section 4.

\section{Mode Identifications and Model Calculations}

In this section, we conducted detailed mode identifications of the DAV star TIC 231277791. Then the input physics and model calculations are introduced.

\subsection{Mode Identifications of TIC 231277791}

The non-radial pulsation modes are characterised by three integer indicators: $k$, $l$, and $m$, which are the radial order, spherical harmonic degree, and azimuth number, respectively (Unno et al. 1989). Brickhill (1975) conducted a research between the rotation period ($P_\mathrm{rot}$) and the corresponding frequency splitting value ($\delta\nu_{k,l}$), and derived an approximate relationship between them, truncated up to the first order of rotational perturbation, as:
\begin{equation}
m\delta\nu_{k,l}=\nu_{k,l,m}-\nu_{k,l,0} \sim \frac{m}{P_\mathrm{rot}}(1-\frac{1}{l(l+1)}),
\end{equation}
where $m$\,$\in$\,$[-\ell, +\ell]$. The observed triplets correspond to the $l$\,=\,1 modes, while the quintuplets are associated with the $l$\,=\,2 modes. According to Equation\,(1), the relationship between $\delta\nu_{k,1}$ and $\delta\nu_{k,2}$ is
\begin{equation}
\frac{\delta\nu_{k,1}}{\delta\nu_{k,2}} \sim \frac{3}{5}.
\end{equation}
\noindent The above proportional relationship is crucial for mode identifications in WDs.

\begin{table*}
\begin{center}
\caption{Mode identifications of the DAV star TIC 231277791. The observed modes are from Romero et al. (2022). The frequencies, intervals between adjacent frequencies, corresponding periods, and amplitudes are listed in columns 2, 3, 4, and 5, respectively.}
\begin{tabular}{lllllllllllllllll}
\hline
ID             &Freq.           &$\delta$F   &Peri.        &A.          &\multicolumn{2}{c}{Mode identification$\_{1}$}&\multicolumn{2}{c}{Mode identification$\_{2}$} \\
               &($\mu$Hz)       &($\mu$Hz)    &(s)         &(ppt)      &$l$        &$m$        &$l$          &$m$         \\
\hline
$f_{01}$       &$\,$2008.96     &$\,$         &$\,$497.77  &$\,$5.52   &$\,$2      &$\,$0      &$\,$2         &$\,$0       \\
               &$\,$            &$\,$11.56    &$\,$        &$\,$       &$\,$       &$\,$       &$\,$          &$\,$        \\
$f_{02}$       &$\,$1997.40     &$\,$         &$\,$500.65  &$\,$3.64   &$\,$2      &$\,$-2     &$\,$2         &$\,$-2      \\
               &$\,$            &$\,$         &$\,$        &$\,$       &$\,$       &$\,$       &$\,$          &$\,$        \\
$f_{03}$       &$\,$1405.40     &$\,$         &$\,$711.54  &$\,$7.03   &$\,$2      &$\,$0      &$\,$2         &$\,$0       \\
               &$\,$            &$\,$11.56    &$\,$        &$\,$       &$\,$       &$\,$       &$\,$          &$\,$        \\
$f_{04}$       &$\,$1393.84     &$\,$         &$\,$717.44  &$\,$3.33   &$\,$2      &$\,$-2     &$\,$2         &$\,$-2      \\
$f_{05}$       &$\,$1390.05     &$\,$         &$\,$719.40  &$\,$3.20   &$\,$1      &$\,$+1     &$\,$1         &$\,$+1      \\
               &$\,$            &$\,$3.13     &$\,$        &$\,$       &$\,$       &$\,$       &$\,$          &$\,$        \\
$f_{06}$       &$\,$1386.92     &$\,$         &$\,$721.02  &$\,$5.14   &$\,$1      &$\,$0      &$\,$1         &$\,$0       \\
               &$\,$            &$\,$3.30     &$\,$        &$\,$       &$\,$       &$\,$       &$\,$          &$\,$        \\
$f_{07}$       &$\,$1383.62     &$\,$         &$\,$722.74  &$\,$2.85   &$\,$1      &$\,$-1     &$\,$1         &$\,$-1      \\
$f_{08}$       &$\,$1331.81     &$\,$         &$\,$750.86  &$\,$3.29   &$\,$1or2   &$\,$0?     &$\,$2         &$\,$+2      \\
               &$\,$            &$\,$10.56    &$\,$        &$\,$       &$\,$       &$\,$       &$\,$          &$\,$        \\
               &$\,$1321.25?    &$\,$         &$\,$756.86? &$\,$       &$\,$       &$\,$       &$\,$2?        &$\,$0?      \\
               &$\,$            &$\,$10.57    &$\,$        &$\,$       &$\,$       &$\,$       &$\,$          &$\,$        \\
$f_{09}$       &$\,$1310.68     &$\,$         &$\,$762.96  &$\,$3.20   &$\,$1or2   &$\,$0?     &$\,$2         &$\,$-2      \\
$f_{10}$       &$\,$1302.44     &$\,$         &$\,$767.79  &$\,$2.80   &$\,$1or2   &$\,$0?     &$\,$1or2      &$\,$0?      \\

\hline
\end{tabular}
\end{center}
\end{table*}

There are 10 independent modes for TIC 231277791 reported by Romero et al. (2022). The detailed mode identifications are shown in Table 1. The third column is the frequency intervals between adjacent modes, represented as $\delta$F with unit in $\mu$Hz. The $\delta$F value between the modes of $f_{05}$ and $f_{06}$ is 3.13\,$\mu$Hz, and that of $f_{06}$ and $f_{07}$ is 3.30\,$\mu$Hz. Therefore‌, the modes of $f_{05}$, $f_{06}$, and $f_{07}$ can be identified as a complete triplet. The average value between 3.13\,$\mu$Hz and 3.30\,$\mu$Hz is 3.215\,$\mu$Hz ($\delta\nu_{k,1}$). According to Equation\,(2), the value of $\delta\nu_{k,2}$ is 5.358\,$\mu$Hz. The $\delta$F value (11.56\,$\mu$Hz) between the modes of $f_{01}$ and $f_{02}$, or $f_{03}$ and $f_{04}$, is approximately twice of $\delta\nu_{k,2}$. It indicates that the four modes are probably component modes of two incomplete quintuplets, with $\Delta$$m$\,=\,2. We can see that the amplitude of the mode of $f_{05}$, $f_{06}$, and $f_{07}$ is 3.20, 5.14, and 2.85\,ppt in the fifth column respectively. The amplitude of the $m$\,=\,0 mode is roughly 1.6$-$1.8 times larger than that of the $m$\,=\,$\pm$1 modes. Considering the larger amplitudes of the modes of $f_{01}$ and $f_{03}$, we identify them as the $m$\,=\,0 modes and assign $f_{02}$ and $f_{04}$ to the $m$\,=\,-2 modes, respectively, as shown in the last four columns in Table 1. According to Equation\,(1), the rotation period of TIC 231277791 is derived to be 41.64\,$\pm$\,2.73\,h. The rotation period of TIC 231277791 is roughly compatible with previous estimates of other WDs (Hermes et al. 2017; Romero et al. 2022; Bogn$\acute{a}$r \& S$\acute{o}$dor 2024).

In Table 1, we conducted two mode identifications. In the mode identification$\_{1}$ (MID$\_{1}$), we assume the modes of $f_{08}$, $f_{09}$, and $f_{10}$ to be $l$\,=\,1 or 2, $m$\,=\,0 modes. We notice that the frequency interval (21.13\,$\mu$Hz) between $f_{08}$ and $f_{09}$ is approximately four times of $\delta\nu_{k,2}$. In the mode identification$\_{2}$ (MID$\_{2}$), we assume $f_{08}$ and $f_{09}$ as $m$\,=\,+2 and -2 modes of an incomplete quintuplet. The $m$\,=\,0 mode is assumed to be 1321.25\,$\mu$Hz. The mode of $f_{10}$ is also assumed to be a $l$\,=\,1 or 2, $m$\,=\,0 mode. In the MID$\_{1}$, we obtain one $l$\,=\,1, $m$\,=\,0 mode, two $l$\,=\,2, $m$\,=\,0 modes, and three $l$\,=\,1 or 2, $m$\,=\,0 modes. In the MID$\_{2}$, we obtain one $l$\,=\,1, $m$\,=\,0 mode, three $l$\,=\,2, $m$\,=\,0 modes, and one $l$\,=\,1 or 2, $m$\,=\,0 mode. The two mode identifications are used to constrain the fitting models. Additionally, it is worth noting that in MID$\_{1}$, the period intervals among the modes of $f_{06}$, $f_{08}$, $f_{09}$, and $f_{10}$ are very close, indicating the possibility of mode trapping effect (Winget, Van Horn \& Hansen 1981).

\subsection{Input Physics and Model Calculations}

The White Dwarf Evolution Code (\texttt{WDEC}) is a program to study the cooling process of WDs. It takes the abundances of elements H, He, C, and O into account. It was originally developed by Martin Schwarzschild and subsequently undergone multiple revisions, such as Kutter \& Savedoff (1969), Lamb \& Van Horn (1975), and Wood (1990). The updated version of the \texttt{WDEC} was released by Bischoff-Kim \& Montgomery in 2018. The \texttt{WDEC} (2018) adopts the equation of state (EOS) and opacity tables of the Modules for Experiments in Stellar Astrophysics (MESA, Paxton et al. 2011, and their subsequent references) in version r8118. We use \texttt{WDEC} (2018, v16) to evolve the grid of DAV star models. The standard mixing length theory (B$\acute{o}$hm \& Cassinelli 1971) is adopted with a mixing length parameter of $\alpha$\,=\,0.6 (Bergeron et al. 1995). The \texttt{WDEC} (2018, v16) is a quasi-static program, which calculates WD evolutions from roughly 10,000\,K to the $T_{\rm eff}$ we needed. The theoretical pulsation modes are synchronously calculated with the WD models.

\begin{table*}
\begin{center}
\caption{The explored parameter spaces of DAV star models evolved by \texttt{WDEC}.}
\begin{tabular}{lllllllllllllllllll}
\hline
Parameters                           &Initial          &Initial         &Middle          &Fine             &{Optimal model$\_{1}$}  &{Optimal model$\_{2}$}          \\
                                     &ranges           &steps           &steps           &steps            &                        &                \\
\hline
$M_\mathrm{*}/M_\mathrm{\odot}$      &[0.500,0.850]     &0.010          &0.005           &0.005            &0.570\,$\pm$\,0.005         &0.720\,$\pm$\,0.005          \\
$T_\mathrm{eff}$ $(K)$               &[10600,12600]     &250            &50              &10               &11300\,$\pm$\,10            &11910\,$\pm$\,10             \\
-log($M_\mathrm{env}/M_\mathrm{*}$)  &[2.00,3.00]       &1.00           &0.05            &0.01             &1.89\,$\pm$\,0.01           &1.90\,$\pm$\,0.01            \\
-log($M_\mathrm{He}/M_\mathrm{*}$)   &[3.00,5.00]       &1.00           &0.05            &0.01             &4.94\,$\pm$\,0.01           &3.09\,$\pm$\,0.01            \\
-log($M_\mathrm{H}/M_\mathrm{*}$)    &[5.00,10.00]      &1.00           &0.05            &0.01             &9.15\,$\pm$\,0.01           &6.11\,$\pm$\,0.01            \\
$X_\mathrm{He}$                      &[0.10,0.90]       &0.16           &0.05            &0.01             &0.32\,$\pm$\,0.01           &0.75\,$\pm$\,0.01            \\
\hline
$X_\mathrm{O}$ of core centre        &               &               &                   &                      &                         \\
\hline
h1                                   &[0.60,0.75]       &0.03           &0.02            &0.01             &0.71\,$\pm$\,0.01           &0.72\,$\pm$\,0.01            \\
h2                                   &[0.65,0.71]       &0.03           &0.02            &0.01             &0.71\,$\pm$\,0.01           &0.68\,$\pm$\,0.01            \\
h3                                   &0.85              &               &0.02            &0.01             &0.82\,$\pm$\,0.01           &0.86\,$\pm$\,0.01            \\
w1                                   &[0.32,0.38]       &0.03           &0.02            &0.01             &0.30\,$\pm$\,0.01           &0.32\,$\pm$\,0.01            \\
w2                                   &[0.42,0.48]       &0.03           &0.02            &0.01             &0.48\,$\pm$\,0.01           &0.50\,$\pm$\,0.01            \\
w3                                   &0.09              &               &0.02            &0.01             &0.10\,$\pm$\,0.01           &0.07\,$\pm$\,0.01            \\
\hline
\end{tabular}
\note {$M_\mathrm{*}$ is the stellar mass, $M_\mathrm{env}$ is the envelope mass, $M_\mathrm{He}$ is the mass of the He layer, and $M_\mathrm{H}$ is the H atmosphere mass. $T_\mathrm{eff}$ $(K)$ represents the effective temperature and $X_\mathrm{He}$ represents He abundance in the mixed C/He region. The O profile in the C/O core is described by six parameters (h1-3, w1-3). The parameter h1 represents the central O abundance, w1 is the mass fraction of $X_\mathrm{O}$\,=\,h1. The parameters h2 and h3 represent the O abundances at the two inflection points of the C/O core. The parameters w2 and w3 are the mass fraction of $X_\mathrm{O}$ from h1 to h2, and from h2 to h3 respectively.}
\end{center}
\end{table*}

The grid parameters of our large sample of DAV star models are shown in Table 2. A total of 7,558,272 DAV star models are constructed using the initial ranges and initial steps. It takes approximately 12\,s for \texttt{WDEC} (2018, v16) to evolve a DAV star model. Using a common four-core computer, we used four terminals for parallel computing. It took almost a year to evolve these large sample of DAV star models. Firstly, the calculated modes are used to fit the observed modes to select an initial optimal fitting model. Subsequently, within a middle step in proximity to the parameters of the initial optimal fitting model, the theoretical pulsation periods are calculated and used to fit the observed modes again, leading to the selection of a middle optimal fitting model. Finally, within a fine step in proximity to the parameters of the middle optimal fitting model, the theoretical pulsation periods are calculated and used to fit the observed modes, until obtaining a final optimal fitting model.

\section{An Asteroseismological Study on TIC 231277791}

In Table 1, we obtained two groups of possible mode identifications. The modes with $m$\,=\,0 in Table 1 are used to constrain the fitting models.
 
\subsection{The Model Fittings}

The evaluation of the fitting results is based on a root-mean-square ($\sigma_{\textup{RMS}}$) equation:
\begin{equation}
\sigma_{\textup{RMS}}=\sqrt{\frac{1}{n} \sum_{n}(P_\mathrm{obs}-P_\mathrm{cal})^2}.
\end{equation}
Here the parameter $P_\mathrm{obs}$ and $P_\mathrm{cal}$ are the observed and calculated periods of the modes, respectively, and $n$ is the number of observed modes. For TIC 231277791, $n$ is 6 in the MID$\_{1}$, and 5 in the MID$\_{2}$. 

We found an initial optimal model among the 7.56 million DAV star models based on the initial parameter space in Table 2. Near the initial optimal model, we adopt the middle step size to evolve the DAV star models. Firstly, we fixed the $X_\mathrm{O}$ parameters of the central core (h1-3, w1-3) and adjusted the six global parameters. Five grid points with the middle steps are adopted for the six global parameters to evolve 15,625 DAV star models. Then, we fixed the six global parameters and adjusted the $X_\mathrm{O}$ parameters of the central core. Five grid points with the middle steps are adopted for the $X_\mathrm{O}$ parameters to evolved 15,625 DAV star models again. We repeated these two steps again and again to obtain a middle optimal model. Similarly, near the middle optimal model, we adopt the fine step size in Table 2 to obtain a final optimal fitting model. More detailed fitting steps can be found in the previous papers (Duan et al. 2021; Chen 2022; Yang et al. 2023; Guo et al. 2023, 2024). After repeated model fittings, we obtained an optimal model$\_{1}$ and an optimal model$\_{2}$, which correspond to MID$\_{1}$ and MID$\_{2}$ in Table 1, respectively. 

The parameters of the optimal model$\_{1}$ are $M_\mathrm{*}$\,=\,0.570\,$\pm$\,0.005\,$M_\mathrm{\odot}$, $T_\mathrm{eff}$\,=\,11300\,$\pm$\,10\,K, -log($M_\mathrm{H}/M_\mathrm{*}$)\,=\,9.15\,$\pm$\,0.01, and -log($M_\mathrm{He}/M_\mathrm{*}$)\,=\,4.94\,$\pm$\,0.01. In the optimal model$\_{2}$, they are $M_\mathrm{*}$\,=\,0.720\,$\pm$\,0.005\,$M_\mathrm{\odot}$, $T_\mathrm{eff}$\,=\,11910\,$\pm$\,10\,K, -log($M_\mathrm{H}/M_\mathrm{*}$)\,=\,6.11\,$\pm$\,0.01, and -log($M_\mathrm{He}/M_\mathrm{*}$)\,=\,3.09\,$\pm$\,0.01. The value of $\sigma_{\textup{RMS}}$ are 0.06\,s and 0.04\,s for the optimal model$\_{1}$ and model$\_{2}$, respectively. The half height and full width of the reciprocal of $\sigma_{\textup{RMS}}$ are used to calculate the errors for each parameter. We notice that when taking the initial step size, middle step size, and fine step size, the fitting error is very close to the corresponding step size. The last two columns in Table 2 show the fitting errors obtained when taking fine step sizes.

In Table 2, we can see that both the values of -log($M_\mathrm{env}/M_\mathrm{*}$) are very close to each other. However, the other five global parameters for optimal model$\_{1}$ are significantly different from those of optimal model$\_{2}$, which indicates the crucial role of mode identifications in model fittings. Taking $X_\mathrm{He}$ as an example, $X_\mathrm{He}$ is 0.32\,$\pm$\,0.01 for the optimal model$\_{1}$ and 0.75\,$\pm$\,0.01 for the optimal model$\_{2}$. Namely, there is a C dominated envelope for the optimal model$\_{1}$ and a He dominated envelope for the optimal model$\_{2}$. The values of the six $X_\mathrm{O}$ parameters for the optimal model$\_{1}$ are similar to those for the optimal model$\_{2}$. According to the optimal model$\_{1}$, TIC~231277791 is likely a relatively small mass and cool DAV star with a slightly thin H atmosphere. While the optimal model$\_{2}$ indicates that TIC~231277791 is likely a relatively massive and hotter DAV star with a slightly thick H atmosphere.

\begin{table*}
\begin{center}
\caption{The table of the detailed fitting results of the optimal model$\_{1}$ and the optimal model$\_{2}$ for TIC 231277791.}
\begin{tabular}{lllllllll}
\hline
&           \multicolumn{2}{c}{Mode identification$\_{1}$}           & &\multicolumn{2}{c}{Mode identification$\_{2}$}       \\
\hline
$P_\mathrm{obs}$   &{$P_\mathrm{cal_{1}}$}  &($l$,$k$)  &{$P_\mathrm{obs}$-$P_\mathrm{cal_{1}}$}   &{$P_\mathrm{cal_{2}}$} &($l$,$k$)&{$P_\mathrm{obs}$-$P_\mathrm{cal_{2}}$} \\ 
(s)         &(s)              &           &(s)                         &(s)             &         &(s)                       \\
\hline
497.77      &$\,$497.74       &(2,12)     &$\,$ 0.03                   &$\,$497.75      &(2,18)   &$\,$ 0.02                 \\
711.54      &$\,$711.54       &(2,19)     &$\,$ 0.00                   &$\,$711.49      &(2,27)   &$\,$ 0.05                 \\
721.02      &$\,$721.05       &(1,10)     &$\,$-0.03                   &$\,$721.02      &(1,15)   &$\,$ 0.00                 \\
750.86      &$\,$750.80       &(2,20)     &$\,$ 0.06                   &$\,$            &         &                          \\
756.86*     &                 &           &                            &$\,$756.91      &(2,29)   &$\,$-0.05                 \\
762.96      &$\,$762.94       &(1,11)     &$\,$ 0.02                   &$\,$            &         &                          \\
767.79      &$\,$767.92       &(2,21)     &$\,$-0.13                   &$\,$767.82      &(1,16)   &$\,$-0.03                 \\
\hline
$\sigma_{\textup{RMS}}$ &              &           &$\,$0.06                 &                &         &$\,$0.04               \\
\hline
\end{tabular}
\end{center}
\end{table*}

In Table 3, we show the detailed fitting results. For the optimal model$\_{2}$, the absolute value of the last column is less than or equal to 0.05\,s, and the value of $\sigma_{\textup{RMS}}$ is 0.04\,s. While, for the optimal model$\_{1}$, there are two larger fitting errors of 0.06\,s and 0.13\,s in the fourth column with $\sigma_{\textup{RMS}}$\,=\,0.06\,s. However, both solutions are of the same order of magnitude. We take the optimal model$\_{2}$ as an example to perform the analysis.

Figure 1 shows the sensitivities for the eight parameters of the optimal model$\_{2}$. In each sub-figure, we adjust the abscissa parameter and fix the other parameters to the values in the last column of Table 2. We can see that the smallest value of $\sigma_{\textup{RMS}}$ for each sub-figure is 0.04\,s. The minimum values in each sub-figure of Fig.\,1 correspond to the values in the last column of Table 2.

\begin{figure*}
\begin{center}
\includegraphics[width=15.0cm,angle=0]{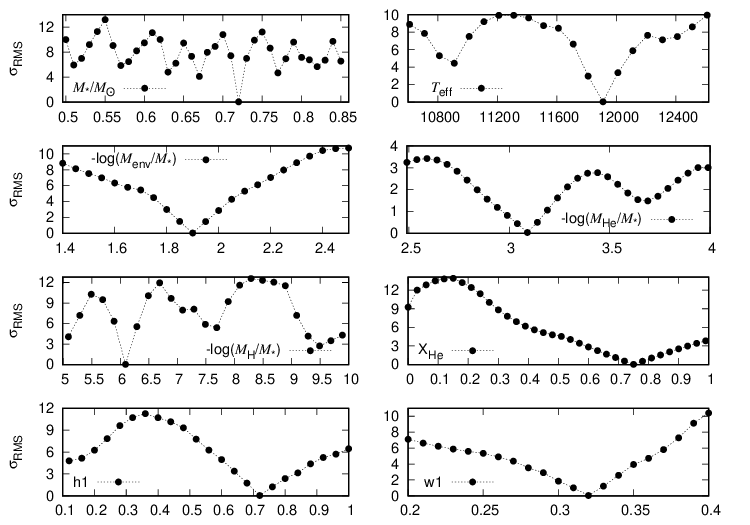}
\end{center}
\caption{Sensitivity figure for eight parameters of the optimal model$\_{2}$. The abscissa in each sub-figure is $M_\mathrm{*}/M_\mathrm{\odot}$, $T_\mathrm{eff}$, -log($M_\mathrm{env}/M_\mathrm{*}$), -log($M_\mathrm{He}/M_\mathrm{*}$), -log($M_\mathrm{H}/M_\mathrm{*}$), $X_\mathrm{He}$, h1, and w1 respectively. The ordinate represents the detailed fitting result $\sigma_{\textup{RMS}}$ for each sub-figure.}
\end{figure*}

Figure 2 shows the fitting error $\sigma_{\textup{RMS}}$ between $M_\mathrm{*}$ and $T_\mathrm{eff}$. The color scale represents the values of $\sigma_{\textup{RMS}}$. Both Fig.\,2(a) (fitting the MID$\_{1}$) and Fig.\,2(b) (fitting the MID$\_{2}$) are the fitting results of 14,271 DAV star models. We can see that the optimal model is around the region (0.570$M_\mathrm{\odot}$, 11,300\,K) in Fig.\,2(a) and around the region (0.720$M_\mathrm{\odot}$, 11,910\,K) in Fig.\,2(b), respectively. The optimal model$\_{1}$ is truly a small mass and cooler DAV star. While, the optimal model$\_{2}$ is truly a massive and hotter DAV star. 

The fitting error $\sigma_{\textup{RMS}}$ between -log($M_\mathrm{H}/M_\mathrm{*}$) and -log($M_\mathrm{He}/M_\mathrm{*}$) is shown in Fig\,3. In Fig.\,3(a), the range of -log($M_\mathrm{H}/M_\mathrm{*}$) is from 6.50 to 10.00 with a step size of 0.01 and the range of -log($M_\mathrm{He}/M_\mathrm{*}$) is from 4.40 to 5.40 with a step size of 0.01. In Fig.\,3(b), the range of -log($M_\mathrm{H}/M_\mathrm{*}$) is from 5.51 to 10.01 with a step size of 0.02 and the range of -log($M_\mathrm{He}/M_\mathrm{*}$) is from 2.61 to 3.51 with a step size of 0.02. There are 35,451 and 10,396 DAV star models in Fig.\,3(a) and Fig.\,3(b) respectively. The optimal model$\_{1}$  has a thin H atmosphere, therefore it is likely to have a strong mode trapping effect.

Figure 4 shows the compositional profiles and the corresponding Brunt-Väisälä frequencies of the optimal model$\_{1}$ and the optimal model$\_{2}$. A WD has a degenerate core surrounded by an outer shell of different chemical elements. In Fig.\,4(b), corresponding the last column in Table 2, we can see that the central O abundance h1 is 0.72, the envelope mass is 1.90 (-log($M_\mathrm{env}/M_\mathrm{*} $)), the He abundance ($X_\mathrm{He}$) is 0.75, the He layer mass (-log($M_\mathrm{He}/M_\mathrm{*}$)) is 3.09, and the H atmosphere (-log($M_\mathrm{H}/M_\mathrm{*}$)) is 6.11. The gradient in the transition zone of the elements in the lower panel leads to the spikes in the upper panel. The number of spikes, as well as their height and width, have a strong influence on the overall structure of the pulsation spectra of stars (Althaus et al. 2010b). In Fig.\,4(a), on close inspection, we can see a small discontinuity around the location of -log(1-$M_\mathrm{r}/M_\mathrm{*} $)\,=\,5. The small discontinuity in He layer is a numerical artifact, which has been reported by previous authors (Bischoff-Kim 2024) and had an effect on the period spectrum of the order of a tenth of a second on the fitness parameter (Kim 2007).

According to the asymptotic theory for $g$-modes, modes (same $l$ and consecutive $k$) have an asymptotic period spacing, while trapped modes have smaller period spacings due to their insertion between normal modes (Brassard et al. 1992). In addition, trapped modes usually have smaller oscillation kinetic energy. In Fig.\,5, we show the period spacing and the oscillation kinetic energy (K.E.) to pulsating periods (Peri.) diagram for the optimal model$\_{1}$. The red dotted lines represent the observed periods in the MID$\_{1}$. The modes of $f_{06}$ and $f_{08}$ are identified as trapped modes in Fig.\,5. For the optimal model$\_{2}$, we have not found any suspected trapped modes. 

\begin{figure*}[htbp]
  \centering
  \begin{subfigure}[b]{0.622\textwidth}
    \includegraphics[width=\textwidth]{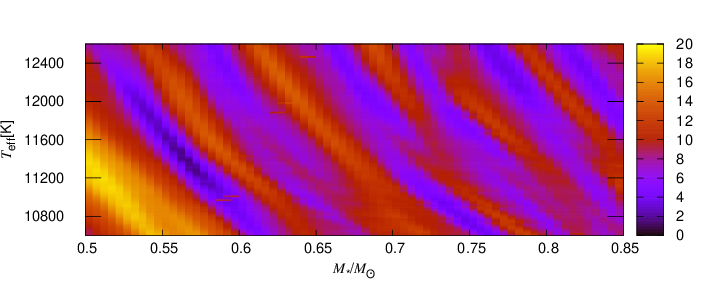}
    \caption{}
  \end{subfigure}
  \begin{subfigure}[b]{0.622\textwidth}
    \includegraphics[width=\textwidth]{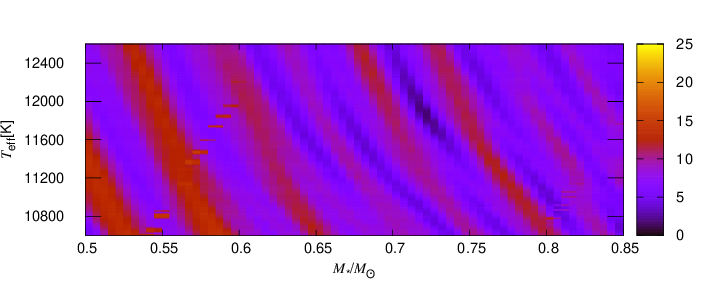}
    \caption{}
  \end{subfigure}
\caption{The color figure between $M_\mathrm{*}$ and $T_\mathrm{eff}$ for the two asteroseismological fitting results. The Fig.\,2(a) is the fitting results for MID$\_{1}$, and the Fig.\,2(b) is the fitting results for MID$\_{2}$. The abscissa spans 0.500 to 0.850\,$M_\mathrm{\odot}$, with a step of 0.005\,$M_\mathrm{\odot}$. The ordinate ranges from 10,600 to 12,600\,K, with a step of 10\,K. Both Fig.\,2(a) and Fig.\,2(b) are fitted by 14,271 DAV star models.
}
\end{figure*}

\begin{figure*}[htbp]
  \centering
  \begin{subfigure}[b]{0.622\textwidth}
    \includegraphics[width=\textwidth]{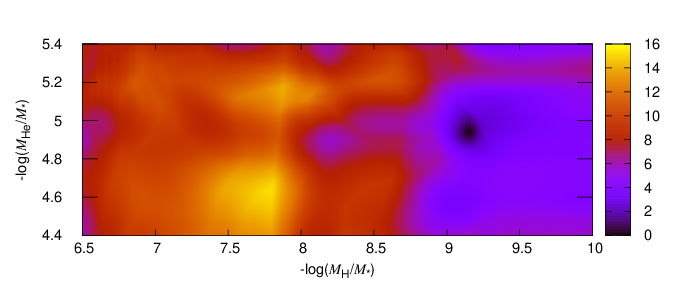}
    \caption{}
  \end{subfigure}
  \begin{subfigure}[b]{0.622\textwidth}
    \includegraphics[width=\textwidth]{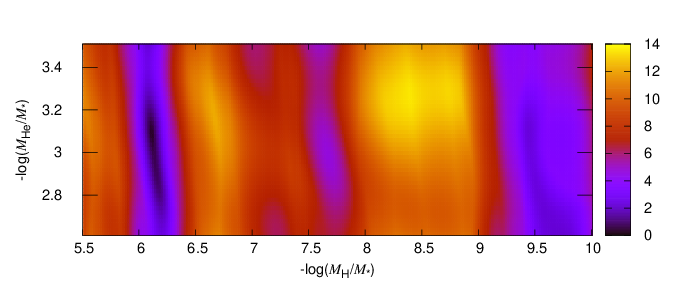}
    \caption{}
  \end{subfigure}
\caption{The color figure between H atmosphere mass and He layer mass for the two asteroseismological fitting results. Fig.\,3(a) and Fig.\,3(b) are fitted by 35,451 and 10,396 DAV star models, respectively.
}
\end{figure*}

\begin{figure*}[htbp]
  \centering
  \begin{subfigure}[b]{0.4975\textwidth}
    \includegraphics[width=\textwidth]{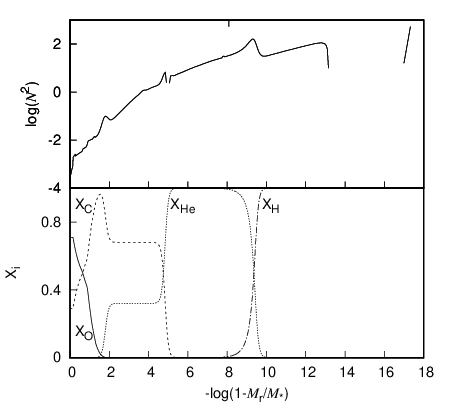}
    \caption{}
  \end{subfigure}
  \begin{subfigure}[b]{0.4975\textwidth}
    \includegraphics[width=\textwidth]{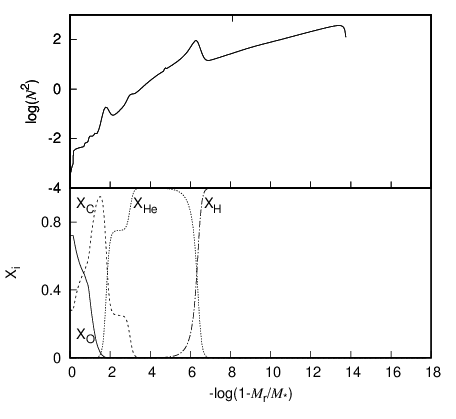}
    \caption{}
  \end{subfigure}
\caption{The core composition profiles and the corresponding Brunt-Väisälä frequency for the two optimal models.
}
\end{figure*}

\begin{figure*}
\begin{center}
\includegraphics[width=14.5cm,angle=0]{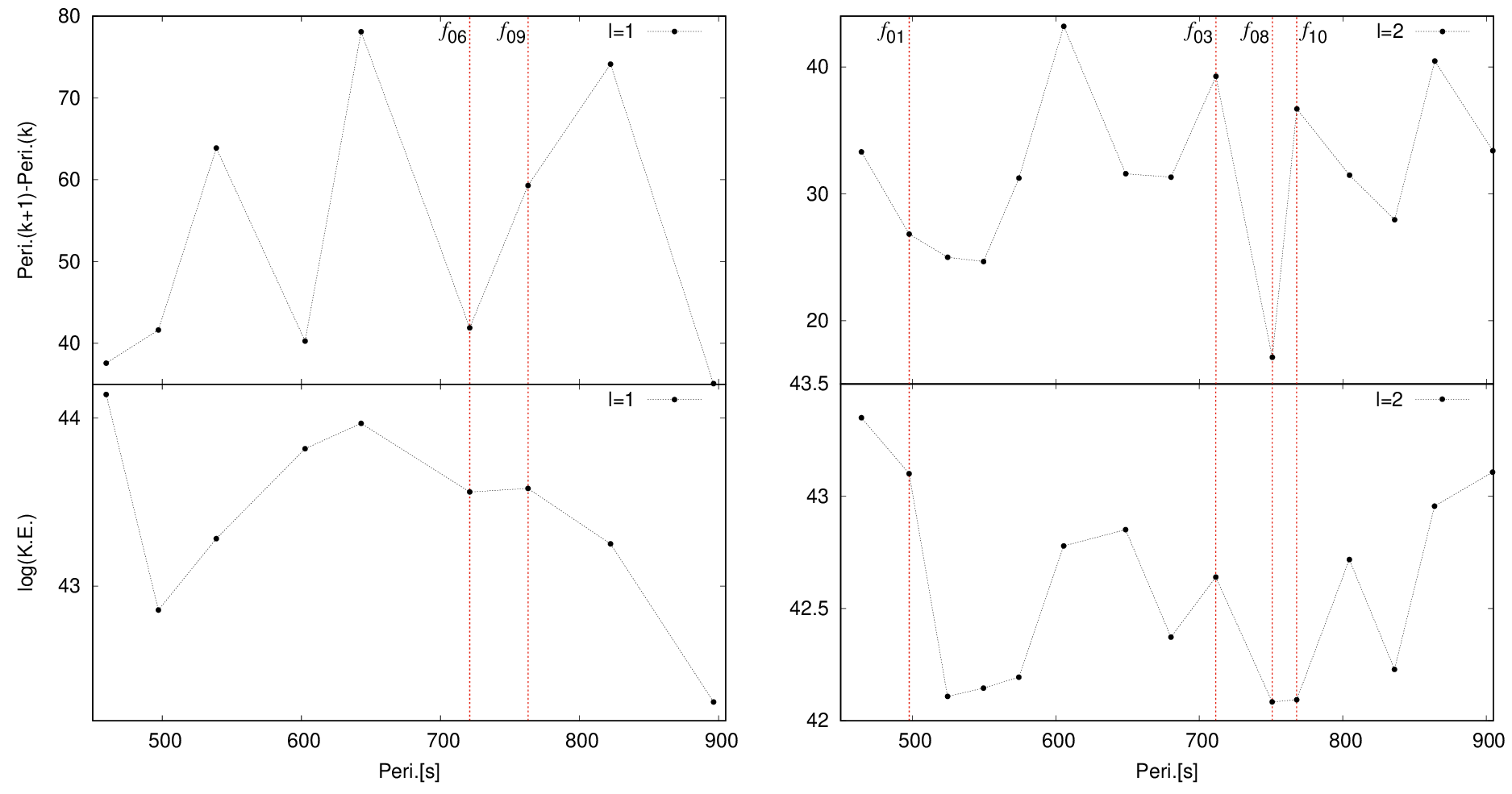}
\end{center}
\caption{The period spacing and the oscillation kinetic energy (K.E.) to pulsating periods (Peri.) diagram for the optimal model$\_{1}$. The dotted lines represent the observed periods for the MID$\_{1}$. The value $k$ is from 5 to 13 for $l$\,=\,1 modes and from 11 to 25 for $l$\,=\,2 modes.
}
\end{figure*}

\subsection{Asteroseismological Distance}

The asteroseismological distance can be calculated using the luminosity of the optimal fitting model (see, e.g., Bogn$\acute{a}$r et al. 2023). We can compare the asteroseismological distance of TIC 231277791 with its Gaia distance. The optimal model$\_{2}$ has a luminosity of log($L/L_\mathrm{\odot}$)\,=\,-2.666. The bolometric magnitude of TIC 231277791 can be calculated using the solar bolometric magnitude $M_\mathrm{bol,\odot}$\,=\,4.74 (Cox 2000) and the modified formula  $M_\mathrm{bol}$\,=\,$M_\mathrm{bol,\odot}$\,--\,2.5$\times$log($L/L_\mathrm{\odot}$) (Torres 2010). The optimal model$\_{2}$ has a $M_\mathrm{bol}$ value of 11.405. The BC values of the DAV star models at $T_\mathrm{eff}$\,=\,11,000\,K and 12,000\,K are -0.441\,mag and -0.611\,mag in V band, respectively (Bergeron et al. 1995). Through difference calculations, we obtain the BC value of -0.596\,mag for the optimal model$\_{2}$ ($T_\mathrm{eff}$\,=\,11,910\,K). The absolute visual magnitude of the optimal model$\_{2}$ is calculated as $M_\mathrm{v}$\,=\,$M_\mathrm{bol}$\,--\,BC(V)\,=\,11.405\,--\,(-0.596)\,=\,12.001. The visual magnitude of TIC 231277791 is $m_\mathrm{v}$\,=\,16.406\,mag (Gaia Collaboration 2023). Using the formula $m_\mathrm{v}$\,--\,$M_\mathrm{v}$\,=\,5log$d$\,--\,5, we can calculate the distance of the optimal model$\_{2}$ to be 76.04\,$\pm$\,0.09\,pc (parallax 13.15\,$\pm$\,0.02\,mas). The calculation of distance error comes from the luminosity error caused by the parameter error in the last column of Table 2. The distance of TIC 231277791 from Gaia is 81.19\,$\pm$\,0.26\,pc (parallax 12.32\,$\pm$\,0.04\,mas, Gaia Collaboration 2022). The distance of the optimal model$\_{2}$ differs from that Gaia by 6.3\%.

The optimal model$\_{1}$ has a luminosity of log($L/L_\mathrm{\odot}$)\,=\,-2.622 and an effective temperature of $T_\mathrm{eff}$\,=\,11,300\,K. We used the same method to obtained an asteroseismological distance of 83.91\,$\pm$\,0.48\,pc (parallax 11.92\,$\pm$\,0.07\,mas) for the optimal model$\_{1}$, which is 3.4\% different from the Gaia distance. For the DAV star TIC 231277791, Vincent et al. (2024) reported a luminosity of log($L/L_\mathrm{\odot}$)\,=\,-2.626, and an effective temperature of $T_\mathrm{eff}$\,=\,11,466\,K. We obtained the corresponding distance of 77.78\,pc (parallax 12.86\,mas), which is 4.2\% different from the Gaia distance. 

We compute the distances of the optimal models and compare them with the distance obtained from Gaia data. The optimal model$\_{1}$ matches the Gaia distance better. Comparing to that from model$\_{1}$, the optimal model$\_{2}$ gives a closer distance than Gaia since more massive stars are smaller and therefore need to be closer to Earth.

\subsection{Comparison with Previous Fitting and Observation Results}

\begin{table*}
\begin{center}
\caption{Table of optimal fitting models. The ID numbers 1, 2, 3, 4, 5 are the spectral results of Jiménez-Esteban et al. (2018), Gentile Fusillo et al. (2019), Gentile Fusillo et al. (2021), Jiménez-Esteban et al. (2023), and Vincent et al. (2024) respectively. The ID numbers 6, 7, and 8 are the asteroseismological results of Romero et al. (2022, 2017, 2024), the optimal model$\_{1}$ of this work and the optimal model$\_{2}$ of this work, respectively.}
\begin{tabular}{llllllllllllllll}
\hline
ID      &$M_\mathrm{*}$       &$T_\mathrm{eff}$  &log\,$g$            &-log($M_\mathrm{H}/M_\mathrm{*}$) &-log($M_\mathrm{He}/M_\mathrm{*}$) & $\sigma_{\textup{RMS}}$\\
        &($M_\mathrm{\odot}$) &(K)               &                    &                                  &                         &(s)           \\
\hline
1       &0.633\,$\pm$\,0.054  &11750             &8.050\,$\pm$\,0.031 &                                  &                         &              \\
2       &0.639  	          &11597             &8.058   	          &                                  &                         &              \\
3       &0.653                &11623\,$\pm$\,235 &8.080\,$\pm$\,0.035 &                                  &                         &              \\
4       &0.612\,$\pm$\,0.018  &11275\,$\pm$\,210 &8.023\,$\pm$\,0.031 &                                  &                         &              \\
5       &0.643 	              &11466      	     &8.065 	          &                                  &                         &              \\
6       &0.570                &11300             &                    &5.45                              &$\sim$1.46               &0.88          \\
7       &0.570\,$\pm$\,0.005  &11300\,$\pm$\,10  &7.975               &9.15\,$\pm$\,0.01                 &4.94\,$\pm$\,0.01        &0.06          \\
8       &0.720\,$\pm$\,0.005  &11910\,$\pm$\,10  &8.212               &6.11\,$\pm$\,0.01                 &3.09\,$\pm$\,0.01        &0.04          \\
\hline
\end{tabular}
\end{center}
\end{table*}

In Table 4, we list the previous spectral and asteroseismological work of TIC 2331277791, as well as the optimal fitting models for the present work. Jiménez-Esteban et al. (2018) (ID\,1) used Gaia Data Release 2 to identify 73,221 WDs within 100\,pc and obtained $M_\mathrm{*}$\,=\,0.633\,$\pm$\,0.054\,$M_\mathrm{\odot}$, $T_\mathrm{eff}$\,=\,11,750\,K and log\,$g$\,=\,8.050\,$\pm$\,0.031 for TIC 231277791. Gentile Fusillo et al. (2019, 2021) (ID\,2, \,3) reported 260,000 and 359,000 high-confidence WDs and derived the stellar parameters for TIC 231277791. Using Gaia DR3, Jiménez-Esteban et al. (2023) (ID\,4) constructed a complete sample of 12,718 WDs within 100\,pc and obtained the stellar parameters for TIC 231277791, Vincent et al. (2024) (ID\,5) created a catalogue of 100,000 high-quality WDs and measured the stellar parameters for TIC 231277791. For ID\,6, Romero et al. (2022) used the fully evolutionary models to perform an asteroseismological analysis for TIC 231277791. The parameters of their best fitting model are $M_\mathrm{*}$\,=\,0.570\,$M_\mathrm{\odot}$, $T_\mathrm{eff}$\,=\,11300\,K, -log($M_\mathrm{H}/M_\mathrm{*}$)\,=\,5.45, and {$\sigma_{\textup{RMS}}$}\,=\,0.88\,s. The ID\,7 and 8 are our optimal model$\_{1}$ and optimal model$\_{2}$ respectively.

The stellar mass for the spectra work for ID\,1 to 5 ranges from 0.612 to 0.653\,$M_\mathrm{\odot}$. The range of $T_\mathrm{eff}$ for ID\,1 to 5 is from 11275 to 11750\,K. The values $M_\mathrm{*}$ and $T_\mathrm{eff}$ of the model ID\,7 are exactly the same as that of the model ID\,6. The values $M_\mathrm{*}$ of ID\,6 and 7 are slightly smaller than that of ID\,1 to 5 and the value $M_\mathrm{*}$ of ID\,8 is slightly larger than that of ID\,1 to 5. The $T_\mathrm{eff}$ of ID\,6 and 7 are basically at the red end of that of ID\,1 to 5 and the value $T_\mathrm{eff}$ of ID\,8 is basically at the blue end of that of ID\,1 to 5. There is a thin H atmosphere for the model ID\,7 and a relatively thick H atmosphere for the models ID\,6 and 8. We can see that the fitting errors of ID\,7 and 8 are reduced by 93\% compared to that of ID\,6 at least. 

\section{A Discussion and conclusions}

This study focuses on the asteroseismological analysis of the DAV star TIC 231277791. Based on 10 detected independent modes reported by Romero et al. (2022), we performed a detailed mode identifications and obtained two relatively reliable mode identifications (MID$\_{1}$ and MID$\_{2}$). The rotation period of 41.64\,$\pm$\,2.73\,h was derived for TIC 231277791 according to the identified triplets and quintuplets. We used the white dwarf evolution code \texttt{WDEC} (2018, v16) to evolve a large sample of DAV star models (7,558,272). The theoretical modes were calculated and used to fit the observed modes of MID$\_{1}$ and MID$\_{2}$. The explored grid parameters of DAV star models and the obtained two optimal fitting models are shown in Table 2. 

For the optimal model$\_{1}$, the parameters are $M_\mathrm{*}$\,=\,0.570\,$\pm$\,0.005\,$M_\mathrm{\odot}$, $T_\mathrm{eff}$\,=\,11300\,$\pm$\,10\,K, -log($M_\mathrm{H}/M_\mathrm{*}$)\,=\,9.15\,$\pm$\,0.01, -log($M_\mathrm{He}/M_\mathrm{*}$)\,=\,4.94\,$\pm$\,0.01, and $\sigma_{\textup{RMS}}$\,=\,0.06\,s. The H atmosphere of the optimal model$\_{1}$ is very thin and two observed modes ($f_{06}$ and $f_{08}$) are identified as trapped modes according to the optimal model$\_{1}$. Based on the optimal model$\_{1}$, the calculated asteroseismological distance is 83.91\,$\pm$\,0.48\,pc, slightly farther than the Gaia distance (81.19\,$\pm$\,0.26\,pc). For the optimal model$\_{2}$, the parameters are $M_\mathrm{*}$\,=\,0.720\,$\pm$\,0.005\,$M_\mathrm{\odot}$, $T_\mathrm{eff}$\,=\,11910\,$\pm$\,10\,K, -log($M_\mathrm{H}/M_\mathrm{*}$)\,=\,6.11\,$\pm$\,0.01, -log($M_\mathrm{He}/M_\mathrm{*}$)\,=\,3.09\,$\pm$\,0.01, and $\sigma_{\textup{RMS}}$\,=\,0.04\,s. The calculated asteroseismological distance, based on the optimal model$\_{2}$, is 76.04\,$\pm$\,0.09\,pc, slightly closer than the Gaia distance. The fitting errors of our optimal model$\_{1}$ and optimal model$\_{2}$ are reduced by 93\% compared to that of Romero et al. (2022) at least.

The central O abundances of optimal model$\_{1}$ and optimal model$\_{2}$ in Table 2 are 0.71 and 0.72, respectively, which are consistent with the results of existing theories of stellar structure and evolution (Salaris et al. 2010 and Althaus et al. 2010b). WDs are the final stage of stellar evolution, and the abundance of central O is influenced by the structure and evolution of its progenitor star. The asteroseismological study on WDs has an opportunity to explore more physical information about their progenitor stars. We need to conduct asteroseismological research on more WDs from the ongoing TESS and upcoming PLATO mission (Rauer et al. 2014). 

\section{Acknowledgment}

The work is supported by the International Centre of Supernovae, Yunnan Key Laboratory (No. 202302AN36000101). The work is also supported by the Yunnan Province Youth Talent Project (2019-182).

\label{lastpage}
\end{document}